\documentclass{moriond}

\def\ra{\rightarrow}

\def\be{\begin{equation}}
\def\ee{\end{equation}}
\def\bea{\begin{eqnarray}}
\def\eea{\end{eqnarray}}

\usepackage{ifthen}
\newboolean{uprightparticles}
\setboolean{uprightparticles}{false} 

\usepackage{acronym}
\usepackage{amsmath,amssymb}
\usepackage{centernot}
\usepackage[super,compress]{cite}
\usepackage{graphicx}
\usepackage{hyperref}
\usepackage{siunitx}
\usepackage{subfigure}
\usepackage{tikz,pgfplots}
\usepackage{xspace}
\usepackage{xcolor}

\usepackage{lineno}

\newcommand{\Photo}{}

\usepackage{bigints}

\usepackage{upgreek}

\def\babar  {\mbox{BaBar}\xspace}
\def\belle  {\mbox{Belle}\xspace}

\def\MagUp {\mbox{\em Mag\kern -0.05em Up}\xspace}

\ifthenelse{\boolean{uprightparticles}}%
{

 \def\Pmu         {\ensuremath{\upmu}\xspace}

 \def\Ppsi        {\ensuremath{\uppsi}\xspace}

 \def\PDelta      {\ensuremath{\Delta}\xspace}                 
 \def\PXi      {\ensuremath{\Xi}\xspace}                 
 \def\PLambda      {\ensuremath{\Lambda}\xspace}                 
 \def\PSigma      {\ensuremath{\Sigma}\xspace}                 
 \def\POmega      {\ensuremath{\Omega}\xspace}                 
 \def\PUpsilon      {\ensuremath{\Upsilon}\xspace}                 
 
 \def\PB      {\ensuremath{\mathrm{B}}\xspace}                 
                  
 \def\PD      {\ensuremath{\mathrm{D}}\xspace}

 \def\PJ      {\ensuremath{\mathrm{J}}\xspace}                 
 \def\PK      {\ensuremath{\mathrm{K}}\xspace}

 \def\Pb      {\ensuremath{\mathrm{b}}\xspace}                 
 \def\Pc      {\ensuremath{\mathrm{c}}\xspace}                 
                  
 \def\Pe      {\ensuremath{\mathrm{e}}\xspace}

 \def\Pi      {\ensuremath{\mathrm{i}}\xspace}

 \def\Ps      {\ensuremath{\mathrm{s}}\xspace}

}
{

 \def\Pmu         {\ensuremath{\mu}\xspace}

 \def\Ppsi        {\ensuremath{\psi}\xspace}                 
                  
 \mathchardef\PDelta="7101
 \mathchardef\PXi="7104
 \mathchardef\PLambda="7103
 \mathchardef\PSigma="7106
 \mathchardef\POmega="710A
 \mathchardef\PUpsilon="7107
                  
 \def\PB      {\ensuremath{B}\xspace}                 
                  
 \def\PD      {\ensuremath{D}\xspace}

 \def\PJ      {\ensuremath{J}\xspace}                 
 \def\PK      {\ensuremath{K}\xspace}

 \def\Pb      {\ensuremath{b}\xspace}                 
 \def\Pc      {\ensuremath{c}\xspace}                 
                  
 \def\Pe      {\ensuremath{e}\xspace}

 \def\Pi      {\ensuremath{i}\xspace}

 \def\Ps      {\ensuremath{s}\xspace}

}

\makeatother

\DeclareRobustCommand{\optbar}[1]{\shortstack{{\miniscule (\rule[.5ex]{1.25em}{.18mm})}
  \\ [-.7ex] $#1$}}

\def\en         {{\ensuremath{\Pe^-}}\xspace}   
\def\ep         {{\ensuremath{\Pe^+}}\xspace}

\def\epem       {{\ensuremath{\Pe^+\Pe^-}}\xspace}

\def\mup        {{\ensuremath{\Pmu^+}}\xspace}
\def\mun        {{\ensuremath{\Pmu^-}}\xspace} 

\def\mumu       {{\ensuremath{\Pmu^+\Pmu^-}}\xspace}

\def\ellell     {\ensuremath{\ell^+ \ell^-}\xspace}

\def\squark    {{\ensuremath{\Ps}}\xspace}

\def\cquark    {{\ensuremath{\Pc}}\xspace}

\def\bquark    {{\ensuremath{\Pb}}\xspace}

\def\kaon    {{\ensuremath{\PK}}\xspace}
  \def\Kbar    {{\kern 0.2em\overline{\kern -0.2em \PK}{}}\xspace}

\def\KorKbar    {\kern 0.18em\optbar{\kern -0.18em K}{}\xspace}

\def\Kp      {{\ensuremath{\kaon^+}}\xspace}

\def\Kstarz  {{\ensuremath{\kaon^{*0}}}\xspace}

  \def\Dbar    {{\kern 0.2em\overline{\kern -0.2em \PD}{}}\xspace}

\def\DorDbar    {\kern 0.18em\optbar{\kern -0.18em D}{}\xspace}

\def\B       {{\ensuremath{\PB}}\xspace}
\def\Bbar    {{\ensuremath{\kern 0.18em\overline{\kern -0.18em \PB}{}}}\xspace}

\def\BorBbar    {\kern 0.18em\optbar{\kern -0.18em B}{}\xspace}
\def\Bz      {{\ensuremath{\B^0}}\xspace}

\def\Bu      {{\ensuremath{\B^+}}\xspace}

\def\Bp      {{\ensuremath{\Bu}}\xspace}

\def\Bd      {{\ensuremath{\B^0}}\xspace}
\def\Bs      {{\ensuremath{\B^0_\squark}}\xspace}

\def\Bc      {{\ensuremath{\B_\cquark^+}}\xspace}

\def\jpsi     {{\ensuremath{{\PJ\mskip -3mu/\mskip -2mu\Ppsi\mskip 2mu}}}\xspace}

  \def\Y#1S{\ensuremath{\PUpsilon{(#1S)}}\xspace}

\def\Lbar        {{\ensuremath{\kern 0.1em\overline{\kern -0.1em\PLambda}}}\xspace}
\def\LorLbar    {\kern 0.18em\optbar{\kern -0.18em \PLambda}{}\xspace}

\def\BF         {{\ensuremath{\mathcal{B}}}\xspace}

\def\BR         {\BF}
\newcommand{\decay}[2]{\ensuremath{#1\!\to #2}\xspace}         
\def\ra                 {\ensuremath{\rightarrow}\xspace}
\def\to                 {\ensuremath{\rightarrow}\xspace}

\def\qsq       {{\ensuremath{q^2}}\xspace}

\def\AT#1     {\ensuremath{A_{\mathrm{T}}^{#1}}\xspace}           

\def\C#1      {\ensuremath{\mathcal{C}_{#1}}\xspace}                       
\def\Cp#1     {\ensuremath{\mathcal{C}_{#1}^{'}}\xspace}                    
\def\Ceff#1   {\ensuremath{\mathcal{C}_{#1}^{\mathrm{(eff)}}}\xspace}        
\def\Cpeff#1  {\ensuremath{\mathcal{C}_{#1}^{'\mathrm{(eff)}}}\xspace}       
\def\Ope#1    {\ensuremath{\mathcal{O}_{#1}}\xspace}                       
\def\Opep#1   {\ensuremath{\mathcal{O}_{#1}^{'}}\xspace}                    

\newcommand{\tev}{\ensuremath{\mathrm{\,Te\kern -0.1em V}}\xspace}
\newcommand{\gev}{\ensuremath{\mathrm{\,Ge\kern -0.1em V}}\xspace}
\newcommand{\mev}{\ensuremath{\mathrm{\,Me\kern -0.1em V}}\xspace}
\newcommand{\kev}{\ensuremath{\mathrm{\,ke\kern -0.1em V}}\xspace}
\newcommand{\ev}{\ensuremath{\mathrm{\,e\kern -0.1em V}}\xspace}
\newcommand{\gevc}{\ensuremath{{\mathrm{\,Ge\kern -0.1em V\!/}c}}\xspace}
\newcommand{\mevc}{\ensuremath{{\mathrm{\,Me\kern -0.1em V\!/}c}}\xspace}
\newcommand{\gevcc}{\ensuremath{{\mathrm{\,Ge\kern -0.1em V\!/}c^2}}\xspace}
\newcommand{\gevgevcccc}{\ensuremath{{\mathrm{\,Ge\kern -0.1em V^2\!/}c^4}}\xspace}
\newcommand{\mevcc}{\ensuremath{{\mathrm{\,Me\kern -0.1em V\!/}c^2}}\xspace}

\def\invfb   {\ensuremath{\mbox{\,fb}^{-1}}\xspace}

\newcommand{\stat}{\ensuremath{\mathrm{\,(stat)}}\xspace}
\newcommand{\syst}{\ensuremath{\mathrm{\,(syst)}}\xspace}

\def\gsim{{~\raise.15em\hbox{$>$}\kern-.85em
          \lower.35em\hbox{$\sim$}~}\xspace}
\def\lsim{{~\raise.15em\hbox{$<$}\kern-.85em
          \lower.35em\hbox{$\sim$}~}\xspace}

\def\tell1  {TELL1\xspace}
\def\ukl1   {UKL1\xspace}

\makeatletter
\def\ja{\@ifstar\@@ja\@ja}
\newcommand{\@ja}[1]{\textcolor{green}{[\textbf{JA:} #1]}}
\newcommand{\@@ja}[1]{\textcolor{green}{#1}}
\def\sr{\@ifstar\@@sr\@sr}
\newcommand{\@sr}[1]{\textcolor{blue}{[\textbf{SR:} #1]}}
\newcommand{\@@sr}[1]{\textcolor{blue}{#1}}
\def\dvd{\@ifstar\@@dvd\@dvd}
\newcommand{\@dvd}[1]{\textcolor{purple}{[\textbf{DvD:} #1]}}
\newcommand{\@@dvd}[1]{\textcolor{purple}{#1}}
\makeatother

\acrodef{BSM}{Beyond the Standard Model}
\acrodef{CKM}{Cabibbo-Kobayashi-Maskawa}
\acrodef{EFT}{Effective Field Theory}
\acrodef{OPE}{Operator Product Expansion}
\acrodef{SM}{Standard Model}

\newcommand{\Rk}{\ensuremath{R_K}\xspace}
\newcommand{\Rkst}{\ensuremath{R_{K^*}}\xspace}

\def \btosll     {\decay{\bquark}{\squark\ellell}}

\def \BtoKstmumu      {\decay{\Bz}{\Kstarz\mup\mun}}

\def \BToKstzll {\decay{\Bz}{\Kstarz\ell^+\ell^-}}

\def \BtoKll {\decay{\Bp}{\kaon^+\ell^+\ell^-}}

\def \BtoKstee      {\decay{\Bz}{\Kstarz\ep\en}}
\def \BtoKmumu      {\decay{\Bp}{\Kp\mup\mun}}
\def \BtoKee      {\decay{\Bp}{\Kp\ep\en}}

\begin{document}
\vspace*{4cm}
\title{Lepton Flavour Universality tests with B decays at LHCb}

\author{ Johannes Albrecht}

\address{Fakult\"at Physik, TU Dortmund, Germany}

\maketitle\abstracts{This article discusses tests of lepton flavour
  universality that are carried out with the LHCb experiment. The
  experimental situation of \btosll and $b \to c \ell \nu$ decays is summarised. }


\section{Introduction}

In the Standard Model of particle physics (SM), the electroweak gauge
bosons $Z^0$ and $W^\pm$ have identical couplings to all three lepton
flavours. This prediction is called lepton flavour universality (LFU)
and is well tested in tree level decays, e.g. of tau leptons, light
mesons or the gauge bosons themselves~\cite{pdg}.

Recent measurements of loop level beauty decays of the type \btosll
and semileptonic beauty decays of the type $b \to c \ell \nu$ have
shown tensions with the SM prediction of LFU. The most precise
measurements of these quantities, performed by the LHCb collaboration,
are summarised in these proceedings. 
All measurements are based on 3\invfb of data
collected at $\sqrt{s}=7\tev$ and 8\tev.

\section{Lepton Flavour Universality in \btosll decays}

A very clean test for new physics can be performed by taking ratios of
\btosll decays to different lepton species. 
At the current experiments, \btosll decays with electrons and muons in the
final state are accessible. If the momentum transfer of the dilepton
system is sufficiently above the dilepton mass, uncertainties in the
hadronic form factors cancel to a very good approximation, leaving a
SM prediction with uncertainties below 1\%~\cite{gino}. 
In the recent years, the interest in lepton flavour universality tests
has increased, mainly driven by two measurements from the LHCb
collaboration: the ratio of \BtoKmumu to \BtoKee, called
\Rk~\cite{Aaij:2014ora}, and the ratio of  \BtoKstmumu to \BtoKstee,
called \Rkst~\cite{Aaij:2017vbb}. The LHCb collaboration uses
basically the same strategy for both analyses, that is discussed here
for general \btosll decays with the corresponding hadron named
$H$. The LFU testing ratio $R_H$ is then
defined as 
\begin{eqnarray*}
R_H = \frac{\bigintssss \frac{ d\Gamma(B \to H \mumu) }{d\qsq} \, d\qsq}{\bigintssss \frac{ d\Gamma(B \to H\epem) }{d\qsq} \, d\qsq} \, ,
\end{eqnarray*}
where the differential decay rate is measured in certain \qsq ranges.
The \qsq ranges corresponding to
the \jpsi and $\psi (2S)$ is always excluded from the LFU analysis and
is used as control channel. To cancel experimental uncertainties in
the absolute efficiencies of the measurements, the ratio $R_H$ is not
measured directly, but as double ratio, normalising the non-resonant
signal mode to the corresponding high-statistics \jpsi mode. The ratio
$R_H$ is then measured as 
\begin{eqnarray*}
R_H = 
{\frac{\BR(\decay{B}{H \mumu})}
{\BR(\decay{B}{H \jpsi(\decay{}{\mumu})})}} 
\bigg{/} 
{\frac{\BR(\decay{B}{H \epem})}
{\BR(\decay{B}{H \jpsi(\decay{}{\epem})})}} \, .
\end{eqnarray*}
A few comments are in order to explain this experimental strategy:
firstly, this method tests for LFU violations in the FCNC decays, it
relies on the conservation of LFU in the corresponding resonant decay
modes. To test this assumption, the ratio of the resonant channels
\begin{eqnarray*}
r(\jpsi) = \frac{\decay{B}{K^{(*)} \jpsi(\ra \mumu)}}
{\decay{B}{K^{(*)}\jpsi(\ra \epem)}}\, ,
\end{eqnarray*}
is confirmed to agree with LFU conservation.
It has to be stressed that this test is a more stringent
test than necessary, because it tests the absolute ratio of muon to
electron reconstruction, identification and selection efficiencies
while in the analyses of $R_H$, only relative efficiencies between
non-resonant and resonant channels are required. If the ratio
$r(\jpsi)$ is tested in bins of the daughter particle momenta, it can
directly test the range of \qsq covered in the analysis. 

The most precise test of $r(\jpsi)$ has been performed in LHCb's
analysis of \Rkst, it was found to be in agreement with unity with a
precision of 4.5\%. Compared to the statistical uncertainties of the
LFU tests of the order of 10\%, this uncertainty is subdominant. For
further tests with enlarged datasets, the precision in the
determination of efficiencies as cross-checked in $r(\jpsi)$ needs to
be studied in more detail. 

\begin{figure}[h]
\begin{minipage}[t]{0.49\linewidth}
\centering
\includegraphics[width=1\linewidth]{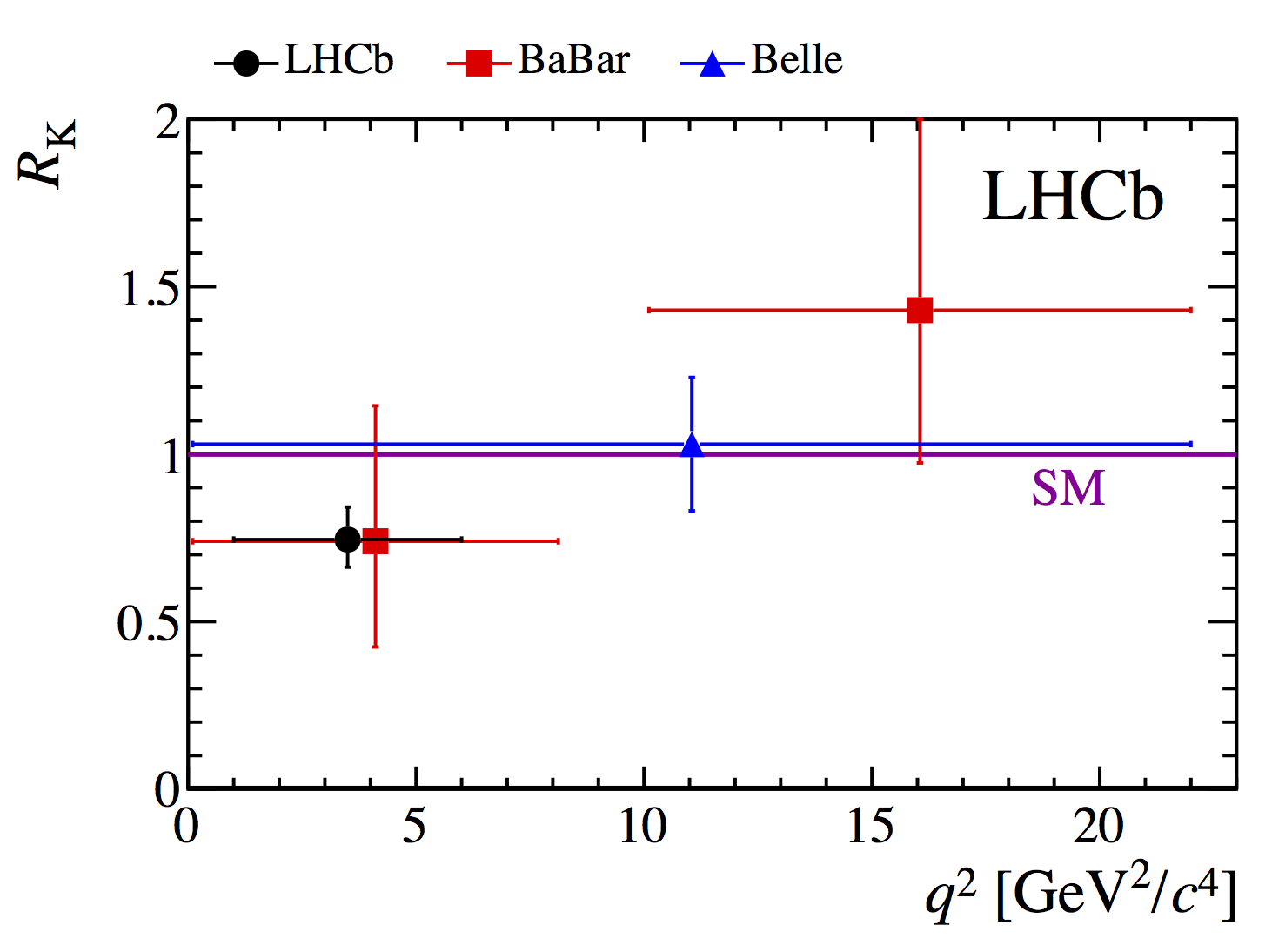}
\end{minipage}
\hspace{\fill}
\begin{minipage}[t]{0.49\linewidth}
\centering
\includegraphics[width=1\linewidth]{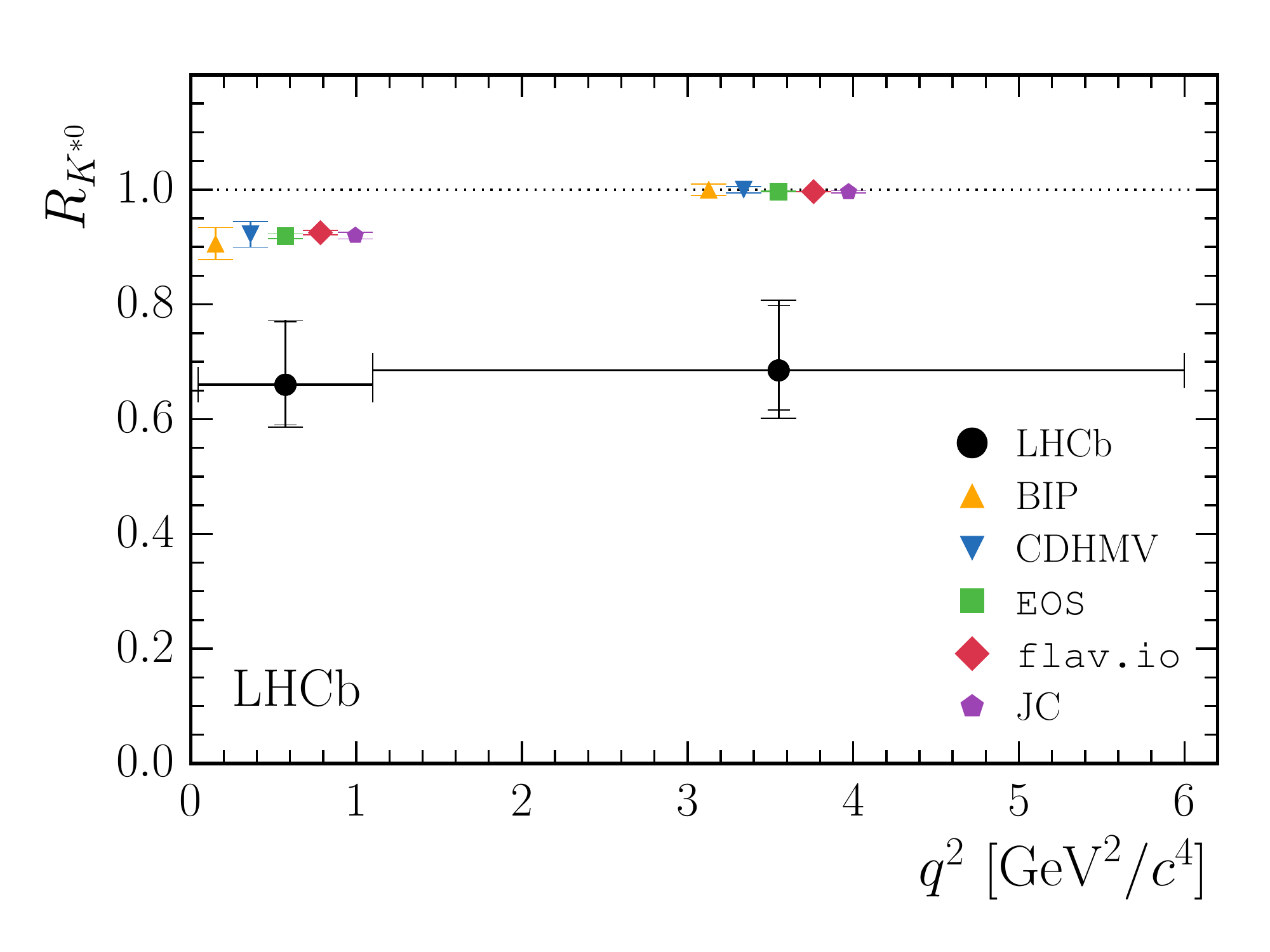}
\end{minipage}
\caption{(left) Summary of the measurements of \Rk of the
  LHCb~\cite{Aaij:2014ora}, \babar~\cite{Lees:2012tva} and
  \belle~\cite{Wei:2009zv} experiments. 
The SM prediction is indicated as a line at 1.0.
(right) LHCb measurement~\cite{Aaij:2017vbb} of \Rkst, together with
several SM predictions. }
 \label{fig:expLFU}
\end{figure}

The experimentally best accessible mode of all \btosll decays is
\BtoKll. The LHCb collaboration published a measurement using 3\invfb
of data~\cite{Aaij:2014ora}. The uncertainty of the measurement is
dominated by the statistical uncertainty of the electron channel, with
a signal yield of $172^{+20}_{-19}$ events, i.e. the statistical
uncertainty is of the order of 12\%. Dominant systematic uncertainties
are the modelling of the mass shape and the determination of the
trigger efficiencies, both accounting for about 3\%. The value of \Rk
is found to be 
\begin{equation}
\Rk=0.745^{+0.090}_{-0.074} {\rm (stat)} \pm 0.036 {\rm (syst)}\, ,
\end{equation}
which is in tension with the SM
prediction~\cite{Bobeth:2007dw} of 1.0 with a significance of 2.6 standard
deviations ($\sigma$). The \babar and \belle experiments have also
published~\cite{Lees:2012tva,Wei:2009zv} tests of LFU,
but their analysed dataset is much smaller than the LHCb dataset and
hence the measurement has significantly larger uncertainties. The
status of all measurements is summarised in
Fig.~\ref{fig:expLFU}~(left).   

The next accessible \btosll channel is \BToKstzll, which has been
published by the LHCb collaboration with 3\invfb with a signal yield
of 89 and 111 events in the low and central bin of $q^2$,
respectively. Similarly to \Rk, the measurement is
implemented as double ratio with the resonant decay mode. Both $q^2$
bins are found below the SM prediction,
\begin{eqnarray*}
\Rkst =
\begin{cases}
0.66~^{+~0.11}_{-~0.07}\stat  \pm 0.03\syst     & \textrm{for } 0.045 < \qsq < 1.1~\gevgevcccc \, , \\
0.69~^{+~0.11}_{-~0.07}\stat \pm 0.05\syst      & \textrm{for } 1.1\phantom{00} < \qsq < 6.0~\gevgevcccc \, .
\end{cases}
\end{eqnarray*}
The measurement of \Rkst is shown in
Fig.~\ref{fig:expLFU}~(right). 
The significances of the deviation of the SM expectation are 2.1 and
2.4~$\sigma$ for the low and middle \qsq bin, respectively. The statistical uncertainty is
of the order of 15\%, dominant systematic uncertainties are due to
data/MC corrections (up to 5\%) and background modelling (up to 5\%).

The LHCb experiment has already collected a factor three more beauty
mesons with respect to the 
3\invfb that are used for the measurements described above. Therefore, the 
tensions seen in the \Rk and \Rkst measurements should get clarified
in the foreseeable future. Then, the \belle~2 experiment will start
to take data and will be able to further test LFU. 

Additionally to the channels discussed above, LFU can be tested in
$\Bs \to \phi \ell^+ \ell^-$ decays, where a first observation of the
channel $\Bs \to \phi e^+ e^-$ should be possible already with 3\invfb
of LHCb data. Also $B^+ \to K^- \pi^+ \pi^- \ell^+ \ell^-$ and
$\Lambda_b \to \Lambda \ell^+ \ell^-$ decays are analysed to test for
violation of lepton universality. Combining the already collected
large datasets and the analysis of more channels, the question if LFU
is conserved in the SM should be conclusively answered in the near
future. A quantitative analysis of the future sensitivities to
discover LFU is discussed in Ref.~\cite{prospects}.

\section{Lepton Flavour Universality in $\decay{\bquark}{\cquark \ell \nu}$ decays}

\begin{figure}[h]
\centering
\includegraphics[width=0.49\linewidth]{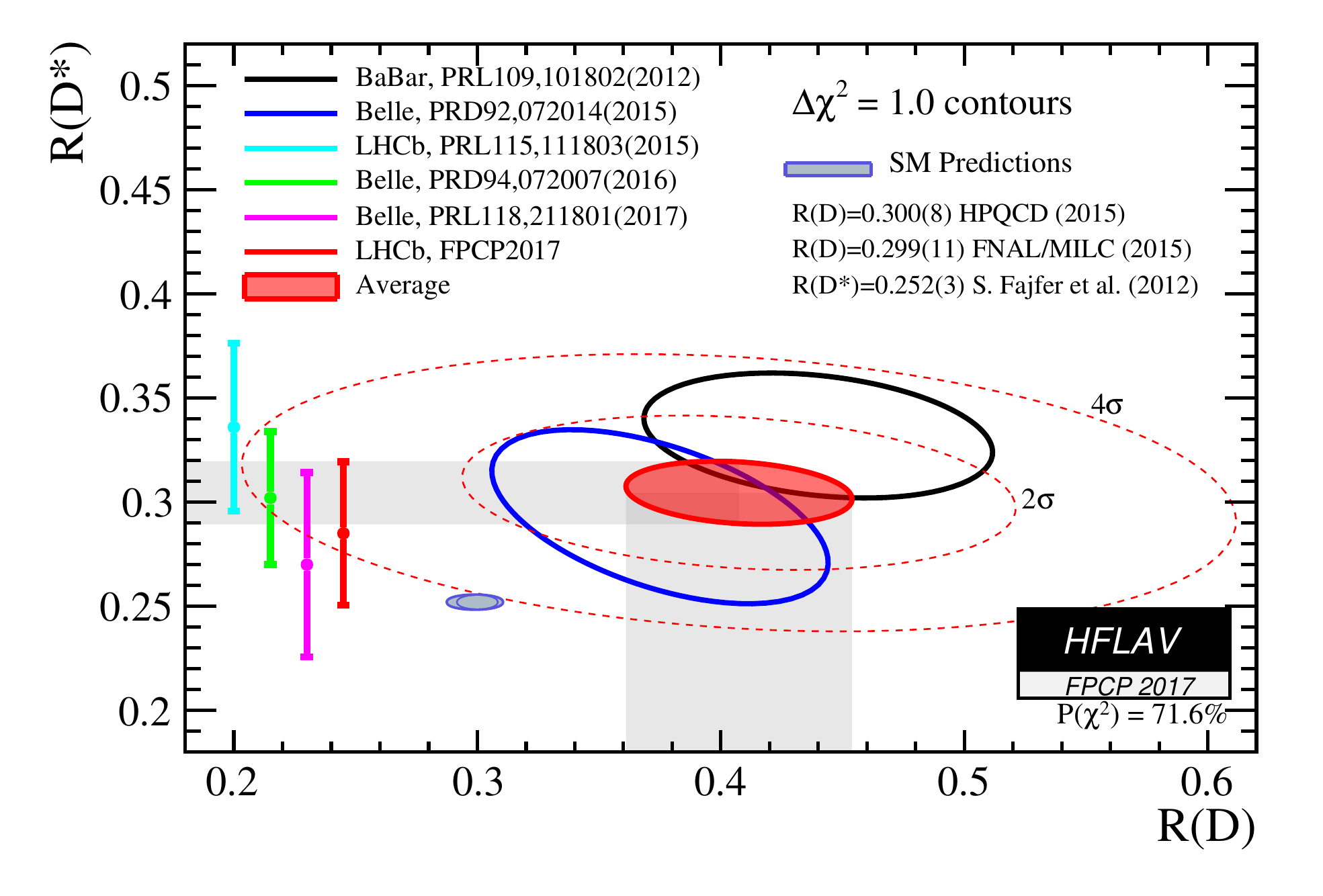}
\caption{Summary of the measurements of $R_D$ and $R_D^*$~\cite{hflav}.}
 \label{fig:expRd}
\end{figure}

Lepton flavour universality can also be tested in semileptonic decays
of the type $\decay{\bquark}{\cquark \ell \nu}$. The observable
$R_D^*$ is defined as \mbox{$R_D^* = \frac{\decay{\bar{\Bd}}{D^{*+} \tau^- \bar{\nu_\tau}}}
{\decay{\bar{\Bd}}{D^{*+} \mu^- \bar{\nu_\mu}}}$.}
The SM prediction is calculated to be \mbox{$R_D^* = 0.252 \pm
0.003$~\cite{rDst}}, the difference to unity originates in the
non-negligible tau-lepton mass. 
LHCb has also measured~\cite{rDstCL21} \mbox{$R_D^* = 0.336 \pm 0.027 \pm 0.030$}, 
using the leptonic $\tau^-$ decay mode \linebreak \mbox{$\tau^- \to \mu^-
  \bar{\nu_\mu}\nu_\tau$}. 
The compatibility with the SM prediction is 2.1$\sigma$. More recently,
LHCb has also measured $R_D^*$ in the hadronic $\tau^-$ decay mode 
$\tau^- \to \pi^+ \pi^- \pi^- (\pi^0) \nu_\tau$, in which the neutral
pion is not reconstructed. $R_D^*$ was measured~\cite{rDstCL22} to a
value of
$0.291\pm0.019\pm0.026\pm0.013$, compatible with the SM prediction at
1.0$\sigma$. The experimental situation of the measurements of $R_D^*$
and also the here not discussed $R_D$ is shown in
Fig.~\ref{fig:expRd}. The combination of both ratios is in tension
with the SM with a significance of 4.1$\sigma$~\cite{hflav}.

LFU can also be tested in \Bc decays. LHCb has performed a measurement
of the ratio $R_\Bc = \BR(\Bc\to\jpsi \tau^+ \nu_\tau) / 
\BR(\Bc\to\jpsi \mu^+ \nu_\mu)$, 
where the $\tau^+$ is reconstructed in the leptonic decay
mode~\cite{rJPsi}. The value found is $R_\Bc =0.71\pm0.17\pm0.18$
which is 2.0$\sigma$ above the SM prediction. 
It should be noted that the tensions seen in $R_\Bc$ are in the same
direction as the tensions seen in $R_{D^{(*)}}$.

\section{Summary}

The recent experimental results testing lepton flavour universality
show intriguing tensions.
In \btosll decays, a tension of $4.0\sigma$ is observed, it is even
increased if the muonic measurements discussed in 
Ref.~\cite{andrew} are included. In $b\to c \ell \nu$ decays, a
tension with a combined significance of also about 4$\sigma$ is
seen. Significant theoretical efforts are ongoing to explain both
types of anomalies in unified models, a detailed discussion can be
found in Refs.~\cite{gudrun,crivelin}. On the experimental side, the
dataset already collected by the LHCb collaboration is about a factor
three larger than the dataset analysed here, so interesting updates on
the presented measurements can be expected in the near future. Also,
the Belle~2 experiment has started to take data and will be able to
provide an important cross check of the measurements from the LHCb
collaboration.

\section*{Acknowledgments}

J. A. gratefully acknowledges support of the Deutsche
Forschungsgemeinschaft (DFG, Emmy Noether programme: AL 1639/1-1) and
of the European Research Council (ERC Starting Grant: PRECISION
714536).



\end{document}